\documentclass[sn-mathphys,iicol]{sn-jnl}
\theoremstyle{thmstyleone}%
\usepackage{caption}
\theoremstyle{thmstyletwo}%

\theoremstyle{thmstylethree}%

\begin{document}

\title[Effect of annealing...]{Effect of annealing in the formation of well crystallized and textured SrFe$_{12}$O$_{19}$ films grown by RF magnetron sputtering}

\author[1]{\fnm{G.D. Soria}}
\equalcont{Current address: Laboratorio Nacional de Fusión, CIEMAT, Madrid, 28040, Spain}

\author[2]{\fnm{A. Serrano}}

\author[1]{\fnm{J.E. Prieto}}

\author[2]{\fnm{A. Quesada}}

\author[3]{\fnm{G. Gorni}}

\author[1]{\fnm{J. de la Figuera}}

\author*[1]{\fnm{J.F. Marco}}\email{jfmarco@iqfr.csic.es}

\affil*[1]{\orgdiv{Instituto de Química Física ‘‘Rocasolano’’}, \orgname{CSIC}, \city{Madrid}, \postcode{28006}, \country{Spain}}

\affil[2]{\orgdiv{Instituto de Cerámica y Vidrio}, \orgname{CSIC}, \postcode{28049}, \country{Spain}}

\affil[3]{\orgdiv{Alba Synchrotron Light Facility}, \orgname{CELLS}, \city{Barcelona}, \postcode{08290}, \country{Spain}}

\abstract{We have studied the influence of annealing treatment on the crystalline growth of SrFe$_{12}$O$_{19}$ previously deposited on Si (100) substrates using radio frequency (RF) magnetron sputtering. For this goal, two grown films, with and without \textit{ex-situ} heating step, have been analysed and compared to determine the differences in their structural, compositional and magnetic properties. The results obtained by the different analysis techniques, in particular Mössbauer spectroscopy together with EXAFS and XANES data, suggest that the as-grown film is composed of nanocrystalline maghemite nanoparticles and amorphous strontium oxide. A strontium hexaferrite canonical structure with c-axis orientation in the sample plane was found for the annealed film.}

\keywords{annealing; crystallization; extended x-ray absorption fine structure (EXAFS); Mösbauer effect; thin film}

\maketitle

\section{Introduction}

\begin{figure}[h!]
\centerline{\includegraphics[width=0.4\textwidth]{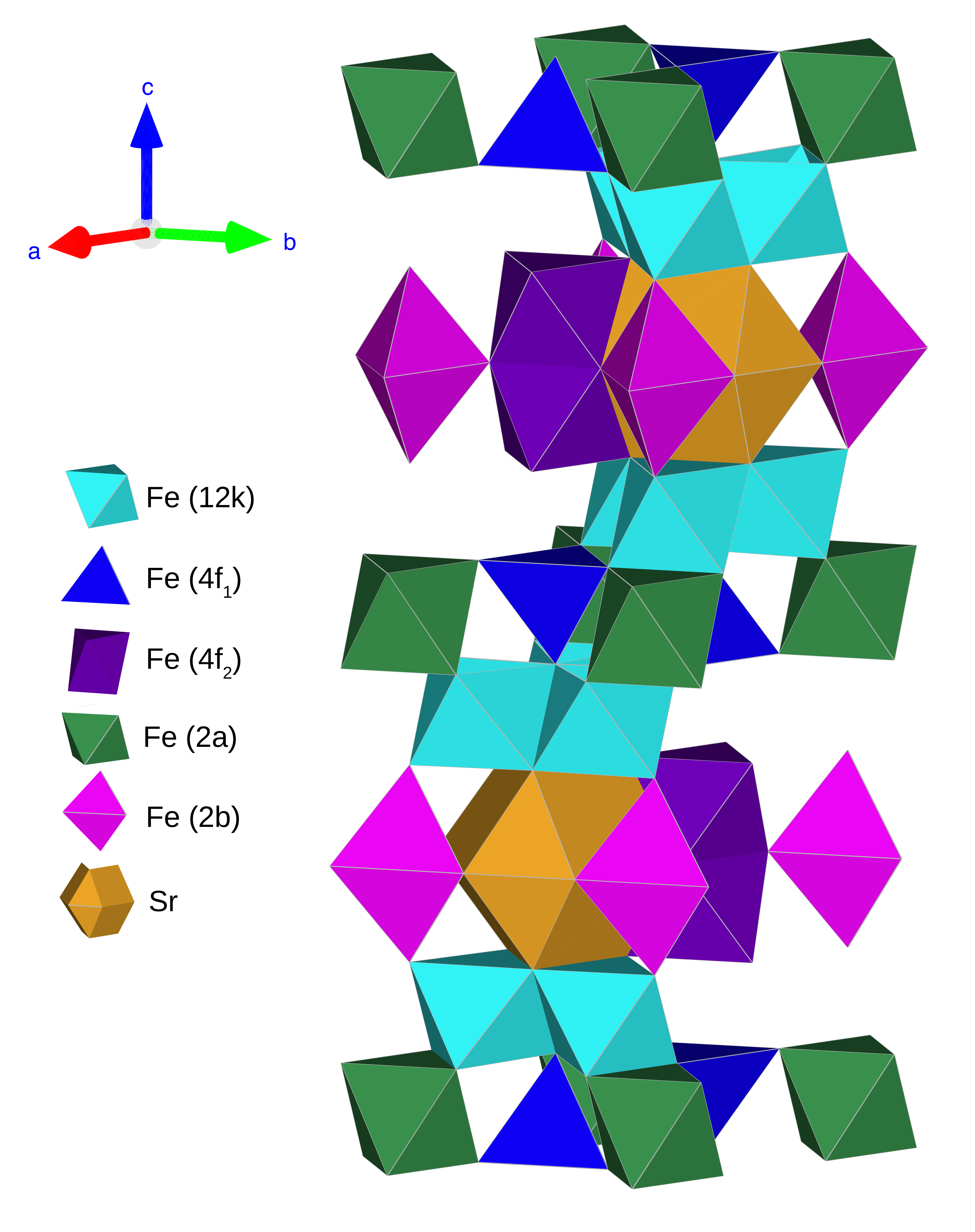}}
\caption{Polyhedral model of the SrFe$_{12}$O$_{19}$ structure. The drawing depicts the chemical environments of iron and strontium in different colours. Wyckoff´s notations are adopted for every iron site. The representation was made using Vesta software \cite{momma_vesta_2011}.}
\label{SFOstructureonly}
\end{figure}

Currently, M-type ferrites are among the most popular permanent magnetic materials. This fact is mainly due to their high magnetocrystalline anisotropy which leads to elevated coercive fields. In addition, these hard ferrites exhibit excellent thermal and chemical stability, high Curie temperatures and good wear resistance, as well as being environmentally friendly and competitively priced products \cite{Brabers, PullarProgMatSci2012}. These materials are widely applied for different uses, such as motors, magneto-mechanical devices and actuators, among others \cite{basak_permanent-magnet_1996, gutfleisch_magnetic_2011, sugimoto_current_2011}. The M-hexaferrite group is composed by strontium hexaferrite SrFe$_{12}$O$_{19}$ (SFO), barium hexaferrite BaFe$_{12}$O$_{19}$ and lead hexaferrite PbFe$_{12}$O$_{19}$ \cite{smit1959ferrites,philips}. Specifically, the hard magnetic material selected in this study has been SFO. The SFO crystal structure (figure \ref{SFOstructureonly}) consists of alternately arranged blocks of spinel and rock salt units. The structure shows a hexagonal closed-packing of oxygen ions, where 2 of the 40 sites are occupied by divalent strontium ions while the trivalent iron ions are located in interstitial sites \cite{smit1959ferrites,kojima_chapter_1982,PullarProgMatSci2012,kreisel_investigation_2001}. Such configuration confers to this material a markedly ferrimagnetic character where iron can be found in five distinct cationic environments: three octahedral sites (12k, 4f2, 2a), one tetrahedral site (4f1) and one bipyramidal site (2b).

In SFO, the easy magnetization axis is the \textit{c} crystallographic axis. Such characteristic of SFO is very promising for both longitudinal and perpendicular recording media applications where magnetic thin films are often required. Strontium hexaferrite films have been grown by pulsed laser deposition \cite{khaleeq-ur-rahman_deposition_2013}, chemical solution deposition \cite{bursik_oriented_2011}, metal-organic chemical vapor deposition \cite{kato_high-speed_2020}, spin coating sol-gel process \cite{masoudpanah_effect_2011} and RF magnetron sputtering \cite{acharya_effect_1996}. For the latter deposition method, it has been reported that the growth of the SFO crystalline phase needs two steps: deposition and post-annealing treatment at a high temperature (800$^{\circ}$C - 900$^{\circ}$C) \cite{eva,acharya_sputter_1994,acharya_effect_1996}. Multiple previous works have established the requirement for the annealing step either \textit{ex-situ} or \textit{in-situ} the deposition chamber  \cite{morisako_effect_1997,ramamurthy_acharya_preparation_1993, hui_effect_2014}. Such studies pointed out by X-ray diffraction (XRD) analysis that the as-grown sample is of amorphous nature. However, only a few studies discuss the character of the as-grown film and how this is affected by the subsequent annealing step. Snyder et al. \cite{snyder_local_1995,snyder_local_1996, snyder_determination_1996} published on the local structural anisotropy in the as-grown film before annealing for barium ferrite thin films by extended X-ray absorption fine structure (EXAFS). The data suggested networks of iron atoms surrounded by their oxygen nearest neighbors, with barium atoms fitting into spaces in-between. Apparently, this determines the final crystalline texture (and the resulting magnetic anisotropy), which forms upon annealing.
Therefore, this study aims to unravel the role of temperature on the aggregation/diffusion of atoms and the consequent formation of the crystalline phase in SFO films deposited by radio frequency (RF) magnetron sputtering. To this end, we have focused on the characterization of a typical as-grown sample and on a sample subject to an annealing treatment after being deposited in identical conditions in the RF magnetron chamber. The characterization includes an ample range of structural, spectroscopic and magnetic techniques such as Rutherford backscattering spectrometry (RBS), XRD, Raman spectroscopy, Mössbauer spectroscopy, vibrating-sample magnetometry (VSM), X-ray absorption near edge structure (XANES) and EXAFS.

\section{Experimental methods}

Two films were deposited by radio frequency (RF) magnetron sputtering of a sintered SrFe$_{12}$O$_{19}$ target made from a commercial SFO powder \cite{commercial}. Silicon (100) wafers 1 mm thick were used as substrates. They were kept at room temperature (RT) during the whole sputtering process. The target-substrate distance was approximately 60 mm and the base pressure in the sputtering chamber was 1 $\times$ 10$^{-6}$ mbar. In order to start from clean targets, a  15 min pre-sputtering process was performed prior to each deposition. A working pressure of 7 $\times$ 10$^{-3}$ of an (Ar/O$_2$) mixture containing 2$\%$ oxygen was maintained inside the sputtering chamber during the deposition process. Samples were grown using a sputtering power of 260 W for five hours. After deposition one of the films was subjected to an annealing treatment at 850 $^{\circ}$C in air for three hours in a separate furnace.

\begin{figure*}[h!]
\centerline{\includegraphics[width=1\textwidth]{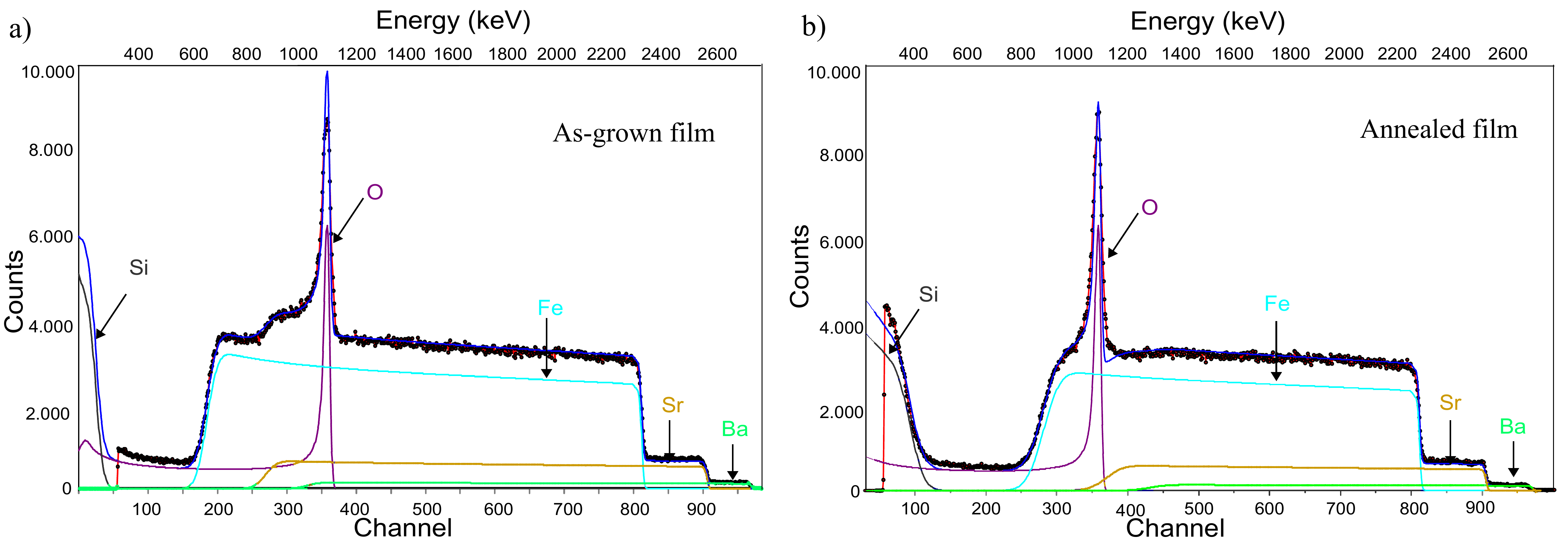}}
\caption{RBS experimental spectra (red circles) of a) an as-grown film and b) an annealed film. The spectrum simulated by the SIMRA code for each sample is given by the blue solid line. The signals of the different elements are identified in each spectrum.}
\label{RBS}
\end{figure*}

The composition and thickness of the different films were determined by RBS using $^{4}$He$^{+}$ at 3.045 MeV to resonantly enhance the O signal. Experiments were performed at the standard beamline of the 5 MV tandem accelerator Centre for Micro-Analysis of Materials (CMAM) in Madrid \cite{redondo-cubero_current_2021}. The ion fluence for the Rutherford backscattering and resonant backscattering experiments was set at 10 $\mu$C. A silicon surface barrier detector, at a scattering angle of 170.0$^{\circ}$, collected the backscattering ions while a three-axis goniometer was employed to control the crystal position.

The crystal structure of the films was examined by X-ray diffraction using a Bruker D8 Advance diffractometer and Cu-K$\alpha$ (1.54 $\r{A}$) radiation in a $\theta$/2$\theta$ configuration. The measuring step was 0.02$^{\circ}$/s with 0.5 s measuring time per step.

Raman spectra were acquired with a micro-Raman Via Renishaw spectrophotometer, equipped with an electrically refrigerated CCD camera, and coupled to a Leica microscope. Radiation of 532 nm from air-cooled Nd: YAG laser was employed for excitation using a diffraction grating of 1800 l/mm. The laser power at the sample was up to 2.0 mW and the spectral resolution was 2 cm$^{-1}$. The acquisition time was 20 sec, the Raman signal was collected over the range 100-1000 cm$^{-1}$ and the objective employed for the measurements was a long-distance 50x Leica.

Mössbauer spectra were collected in the conventional transmission mode using a constant acceleration spectrometer, a ${^{57}}$Co(Rh) source, and a helium closed-cycle cryorefrigerator \cite{GancedoHI1994}. Spectra were recorded at 298 K and 35 K and computer-fitted. The velocity scale was calibrated using a metallic $\alpha$-Fe foil 6 $\mu$m thick. The isomer shifts were referred to the centroid of the spectrum of $\alpha$ Fe at room temperature.

Sr and Fe K-edge XANES and EXAFS data were recorded in the fluorescence mode at the CLAESS beamline of the Alba synchrotron \cite{CLAESS}. Data were collected at 298K and different incident angles of the incoming X-rays (5$^{\circ}$ and 85$^{\circ}$). The energy scale was calibrated using a 6 $\mu$m iron foil and strontium(II) oxide. The position of the first inflection point on the iron foil edge was taken at 7112.3 eV and that of strontium(II) oxide at 16110.2 eV. All the iron-XANES and strontium-XANES data are reported here relative to these values. The edge profiles were separated from the EXAFS data and, after subtraction of a linear pre-edge background, normalized to the edge step. The position of the absorption edge was obtained from the first most intense maximum of the first derivative of the edge profile. The EXAFS oscillations were isolated after background subtraction of the raw data using the software package Athena and converted into \textit{k} space. The data were weighted by \textit{k}$^{2}$, where k is the photoelectron wave vector, to compensate for the diminishing amplitude of EXAFS at high \textit{k}. 

Magnetic properties were studied by recording hysteresis loops at RT with a homemade VSM applying a magnetic field of 1.25 T.

\section{Results and Discussion}

RBS analysis allowed for determining the chemical composition of both samples. The data (figure \ref{RBS}) were fitted using the SIMRA code \cite{mayer_simnra_nodate} obtaining good agreement with the experimental spectra. The as-grown sample and the annealed film presented the same elements in their spectra: O, Si, Fe, Sr and Ba (Fig. \ref{RBS}). Their signals come from the SFO film as well as from the substrate. The barium signal stems from a small Ba impurity in the commercial powder \cite{soria_influence_2020}. The concentrations of the different elements were not significantly different between both films. The atomic Fe/O ratio expected for pure SFO is 0.63. In the case of the non-annealed film, the Fe/O ratio was 0.66, while in the annealed film was 0.63. Thus, it follows that the small percentage of missing oxygen atoms during deposition are recovered after the annealing treatment. In addition, the film thicknesses calculated were 1.86 $\mu$m  and 1.62 $\mu$m  for the as-grown and annealed films, respectively. The observed difference in thickness (13$\%$) is probably due to a deviation in the deposition gradient arising from the samples being placed at a slightly different position in the sample stage of the magnetron device.

\begin{figure}[h!]
\centerline{\includegraphics[width=0.4\textwidth]{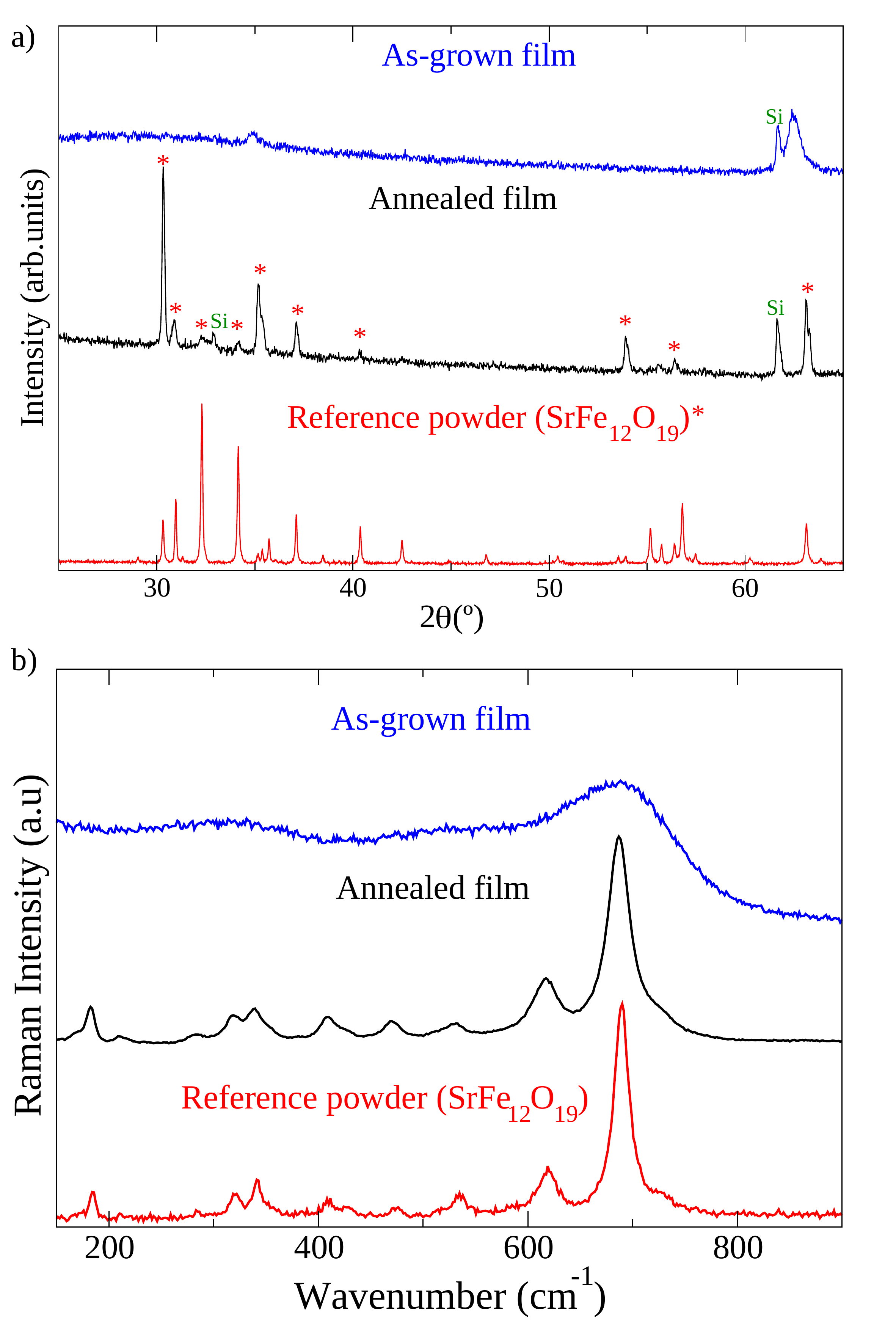}}
\caption{a) XRD diffraction patterns and b) Raman spectra recorded from both films together with the data recorded from a SFO reference powder.}
\label{XRD-Raman}
\end{figure}

The crystal structure, texture and chemical composition of the films were also studied by XRD and Raman spectroscopy. Fig. \ref{XRD-Raman}a shows the XRD patterns recorded from each film together with a reference XRD pattern recorded from a commercial SFO powder. The annealed film shows all the diffraction peaks that correspond to a crystalline strontium hexaferrite structure, besides the Si peaks arising from the silicon substrate. The highest intensity peaks appear at 30.4$^{\circ}$, 35.2$^{\circ}$, 37.1$^{\circ}$, 54.0$^{\circ}$ and 63.2$^{\circ}$ which match with the (110), (201), (203), (300), (220) Miller indices, respectively of SFO \cite{kiani_synthesis_2013,azim_indexing_2014}. The most prominent peak is the one at 30.4$^{\circ}$, i.e, that corresponding to the (110) diffraction plane. This indicates a preferential growth of the SFO structure with the c-axis parallel to the film plane. The crystallite size was determined by the Scherrer equation \cite{scherrer_bestimmung_1918}, and the lattice parameters by the distance formula between adjacent planes in the (hkl) set of the hexagonal crystal structure \cite{cullity_elements_1978}. The calculated average particle size is 55 nm whilst the \textit{a} and \textit{c} lattice parameters are 0.589 nm and 2.349 nm, respectively. The unit cell values are close to those reported in the literature \cite{PullarProgMatSci2012,azim_indexing_2014}. The pattern of the as-grown sample shows only a few broad diffraction peaks, indicating a low crystallinity of the film, which, besides, are not coincidental with those expected for SFO. In fact, the diffraction peaks at 35.1$^{\circ}$ and 66.5$^{\circ}$, are compatible with the presence of the spinel-related phases magnetite (Fe$_{3}$O$_{4}$) or maghemite ($\gamma$-Fe$_{2}$O$_{3}$) \cite{aliahmad_synthesis_2013,ruiz-baltazar_effect_2015,zulfiqar_static_2014, kim_new_2012,chernyshova_size-dependent_2007}. Since XRD does not allow unambiguous identification of these two phases, we resorted to Raman and Mössbauer spectroscopy in order to achieve a more accurate assignment (see below).

\begin{figure*}[h!]
\centerline{\includegraphics[width=0.75\textwidth]{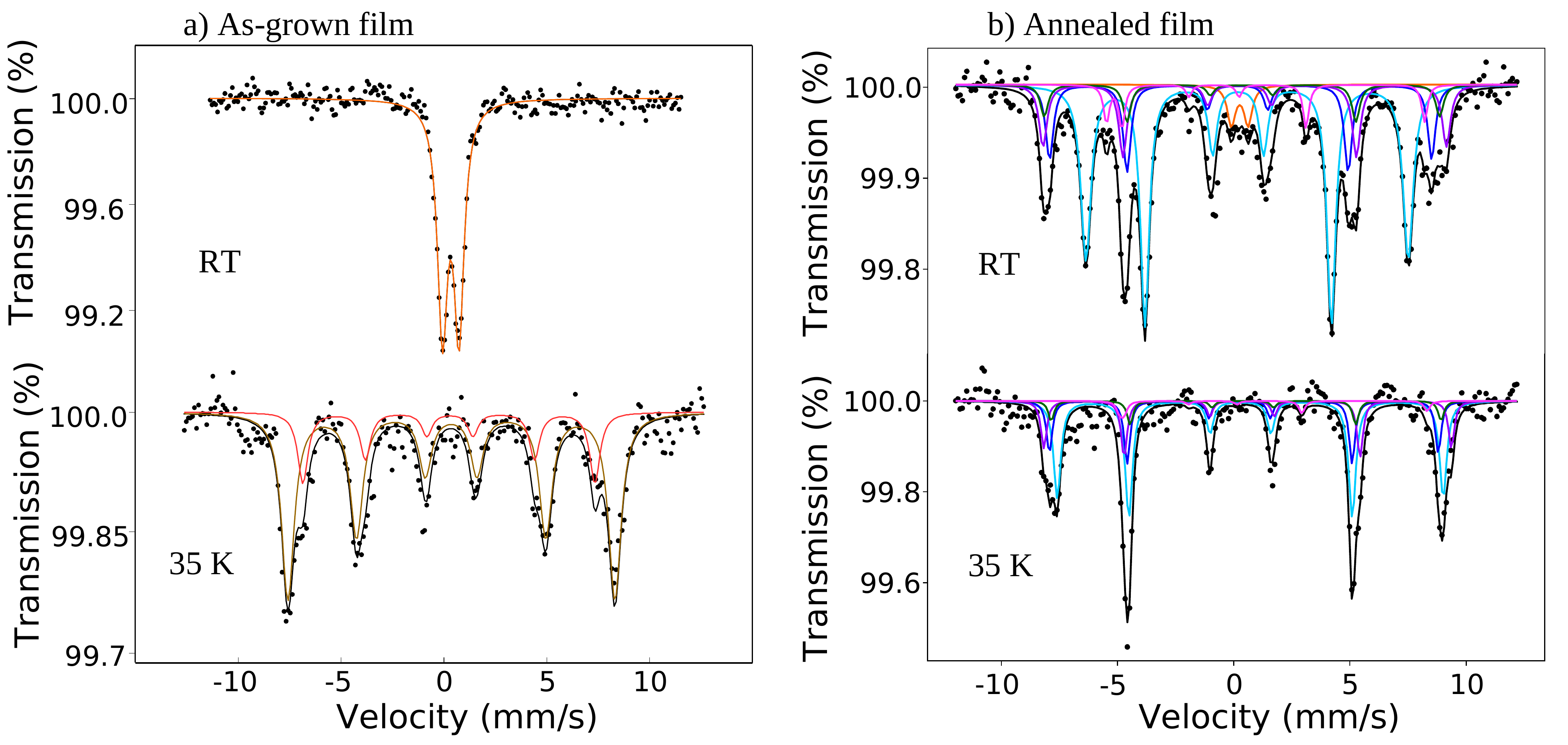}}
\captionsetup{justification=centering}
\caption{Transmission Mössbauer spectra recorded from both films at RT and 35 K.}
\label{SFOmicramossbauerfinal}
\end{figure*}

\begin{table*}[h!]
\centering
\begin{tabular}{|c|c|c|c|c|c|c|}
\hline
T & Site & $\delta$  & 2$\varepsilon$ & H & $\Gamma$ & Area \\
(K) & ($\pm$ 0.03 mms$^{-1}$) & ($\pm$ 0.05 mms$^{-1}$) & ($\pm$ 0.05 T) & ($\pm$ 0.05 mms$^{-1}$) & ($\%$) & \\
\hline
298 & Fe$^{3+}$ & 0.34 & 0.80 & - & 0.40 & 100 \\
35 & Fe$^{3+}_{Td}$ & 0.39 & -0.03 & 44.2 & 0.40 & 28 \\
& Fe$^{3+}_{Oh}$ & 0.46 & 0.01 & 49.3 & 0.40 & 72 \\
\hline
\end{tabular}
\captionsetup{justification=centering}
\caption{${^{57}}$Fe Mössbauer parameters obtained from the fit of the spectra shown in figure \ref{SFOmicramossbauerfinal}a. The symbols $\delta$, $\Delta$, 2$\varepsilon$, H, $\Gamma$ correspond to isomer shift, quadrupole splitting (applies to doublets), quadrupole shift (applies to sextets), hyperfine magnetic field and linewidth, respectively.}
\label{masmaghemite}
\end{table*}

The corresponding Raman spectra are collected in figure \ref{XRD-Raman}b. As in the XRD analysis, a reference SFO powder pattern was used for identification purposes. The spectrum of the annealed sample presents the bands characteristic of strontium hexaferrite \cite{elansary_new_2020,morel_sublattice_2002,kreisel_raman_1999,nethala_investigations_2018,kreisel_raman_1998}. Such Raman modes arise from the iron cations in different crystallographic sites of the SFO structure. The as-grown sample shows a vibration mode with a broad peak at 690-700 cm$^{-1}$. According to the published literature, the broad and most intense peak for maghemite appears around 703 cm$^{-1}$, and can be assigned to the A$_{1g}$ vibration mode (O-Fe-O bridge) \cite{dar_single_2013,santillan_optical_2017,lopez_sensitivity_2009}. This fact correlates well with the XRD data. A presumed amorphous SrO phase was not detected by Raman spectroscopy due to its lower scattering cross-section \cite{SERRANO20221014}.

In figure \ref{SFOmicramossbauerfinal}a the transmission Mössbauer spectra recorded from the as-grown SFO film at 298 K (RT) and 35 K are presented. The RT spectrum does not exhibit the characteristic magnetic hyperfine pattern of SFO \cite{BerryJMSL2001} but a paramagnetic doublet instead. This doublet presents an isomer shift ($\delta$ = 0.34 mm/s) and a quadrupole splitting ($\Delta$ = 0.80 mm/s) which are characteristic of high spin Fe$^{3+}$ \cite{menil_systematic_1985}. The lack of magnetic ordering at RT together with the broad signals shown by both the XRD and Raman data suggests that the as-grown film could be of superparamanetic/low crystallinity nature associated to the reduced size of the film grains. The 35 K spectrum shows two relatively well-separated magnetic contributions whose Mössbauer parameters (Table \ref{masmaghemite}) are close to those shown by Fe$^{3+}$ cations occupying the octahedral and tetrahedral sites of a spinel-related structure. The low-temperature spectrum registered from the as-grown sample does not correspond, then, with that expected for SFO and resembles clearly that shown by many oxides having a spinel-related structure. In this particular case, the 35 K spectrum, and the corresponding hyperfine parameters, are very similar to those reported for 5 nm maghemite nanoparticles measured at comparable temperatures (40 K) \cite{roca_effect_2007}. Both the RT and 35 K Mössbauer spectra endorse the XRD and Raman results and confirm that the iron phase contained in the as-grown film is, most likely, superparamagnetic maghemite. In contrast, the RT and 35 K spectra recorded from the annealed film (figure \ref{SFOmicramossbauerfinal}b) are fully typical of SFO \cite{BerryJMSL2001}, except by the small Fe$^{3+}$ paramagnetic doublet (3$\%$) seen at RT, which possibly arises from a minor fraction of SFO/maghemite particles of nanometer dimensions. 

\begin{figure*}[h!]
\centerline{\includegraphics[width=0.65\textwidth]{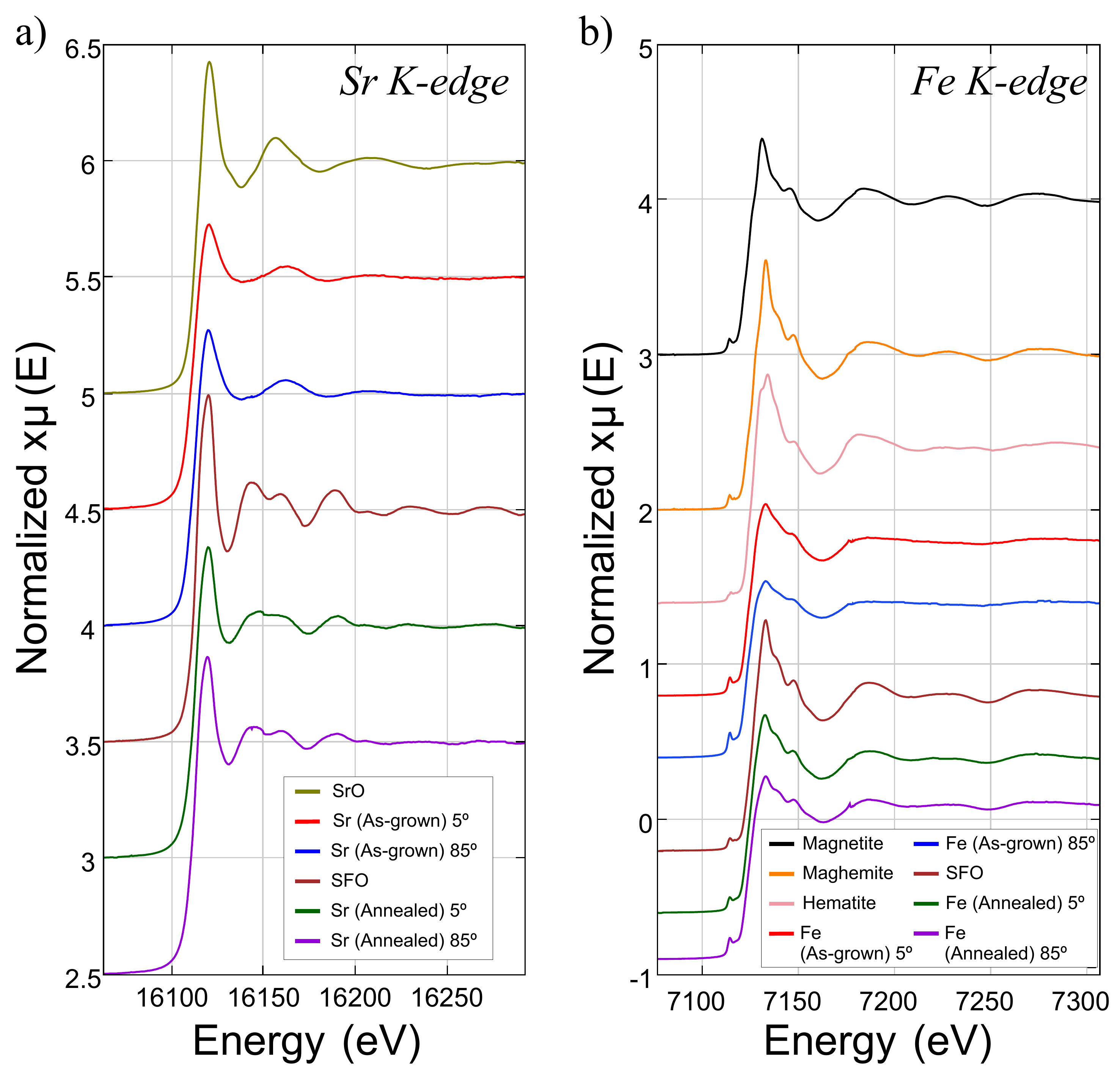}}
\captionsetup{justification=centering}
\caption{a) and b) XANES spectra at the Sr and Fe K-edge, respectively, recorded from the as-grown and annealed films at different setup orientations. The data recorded from different reference compounds are also shown.}
\label{XANES}
\end{figure*}

The intensity ratio of the absorption lines of the sextets (3:x:1:1:x:3) can be used to determine the average orientation of the magnetization in each sample. The spectrum of the as-grown sample (\ref{SFOmicramossbauerfinal}a) at 35 K is best fitted using a value of x = 2, indicating a random orientation of the magnetization. However, for the annealed sample a x = 3.5 value is obtained which corresponds to an average magnetization mostly within the plane (15$^{\circ}$) and indicates that the film is highly magnetically textured.

Since the data so far have indicated that the as-grown film contains maghemite of nanometric dimensions, a pertinent question which is still unanswered refers to the chemical state of strontium in that film. In order to understand this point, we recorded both Sr- and Fe-K edge XANES and EXAFS data. Figure \ref{XANES}a shows the Sr-K edge XANES of the reference compounds SrO and SFO together with those of the as-grown and annealed films at different incident angles of the incoming X-rays (5$^{\circ}$ and 85$^{\circ}$). The two different orientations were used in order to look into the existence of a local structural orientation preference in the films. Inspection of this figure indicates that the position of the edge, defined as the first inflexion point of the edge, i.e., the point where the first derivative has its first maximum, is the same in all the cases (16112.7 eV) and, therefore, indicative of the presence of Sr$^{2+}$ in both films. The Sr K-edge XANES data of the as-grown and the annealed film are clearly different, while those corresponding to the as-grown film resemble that of SrO, and the spectra of the annealed film match the spectrum of SFO. 
As compared with the SrO spectrum, is clear that the oscillations beyond the white line in the spectra of the as-grown film are considerably attenuated what suggests a more disordered character of this sample. This does not occur in the case of the spectra of the annealed film when compared with the reference spectrum of SFO. Finally, it appears that there are no significant differences between the spectra taken at different incident angles of the X-ray beam. This result is contrary to prior studies which pointed out that the anisotropic local structure in the as-grown sample determines the crystalline orientation of the annealed film \cite{snyder_local_1995,snyder_local_1996, snyder_determination_1996}.

\begin{figure*}[h!]
\centerline{\includegraphics[width=0.7\textwidth]{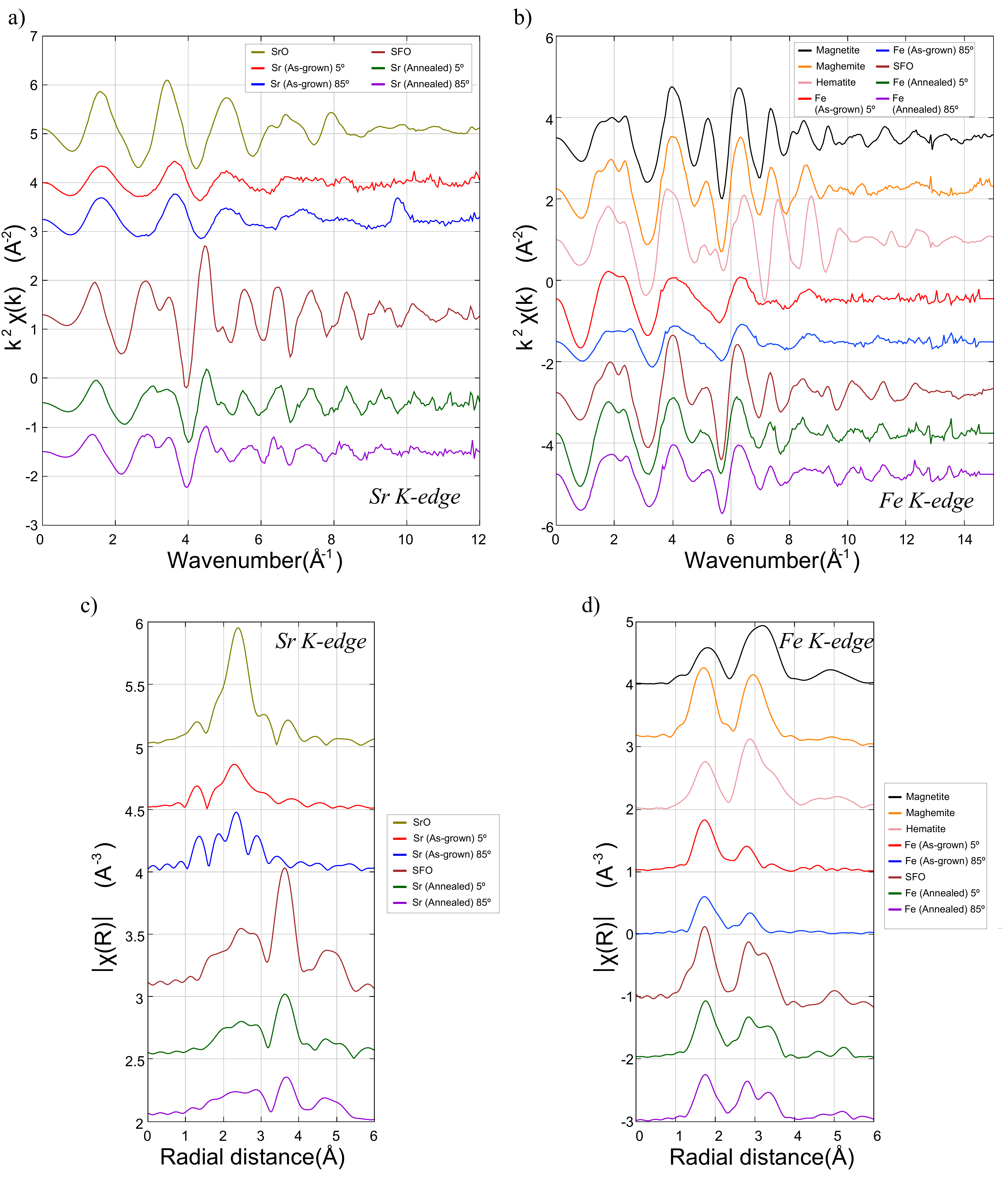}}
\caption{a) and b) EXAFS data at the Sr and Fe K-edge, respectively, recorded from the as-grown and annealed films at different setup orientations. The data recorded from different reference compounds are also shown. c) and d) Fourier-transformed data at the Sr and Fe K-edge, respectively, recorded from the as-grown and annealed films at different setup orientations together with the Fourier-transform data recorded from different reference compounds.}
\label{EXAFS}
\end{figure*}

Figure \ref{XANES}b collects the Fe K-edge XANES recorded from the reference compounds magnetite, maghemite, hematite and SFO together with the spectra recorded (again at two different incidence angles of the X-ray beam) from the as-grown and annealed films. The position of the edge of the spectra recorded from the two films is located at 7123.2 eV, a value which coincides with that of the edge of the spectra of maghemite and hematite, therefore confirming, as Mössbauer spectroscopy had anticipated, the presence of Fe$^{3+}$ in the samples. The position of the edge of magnetite, in which 50$\%$ of the octahedral sites are occupied by Fe$^{2+}$, appears at the lower energy of 7121.0 eV, what is consistent with the mentioned presence of Fe$^{2+}$ in this material. The spectra of the as-grown film show a very intense pre-edge peak at 7114.5 eV indicating the existence of tetrahedral environments in the sample. This endorses the tetrahedral component observed in the 35K Mössbauer spectrum.  Similarly to that observed in the Sr K-edge XANES of this sample, the oscillations beyond the white line are very much attenuated which is consistent with its low crystallinity as evidenced by the rest of the diffraction and spectroscopic techniques. Because of the strong attenuation of the mentioned oscillations, it is hard to associate the Fe K-edge XANES spectrum of the as-grown-film with any of the reference compounds. However, the Fe K-edge XANES data recorded from the annealed film are very similar to that recorded from the reference SFO.

Figure \ref{EXAFS}a shows the Sr K-edge EXAFS recorded from SrO, SFO, the as-grown film and the annealed film. The data recorded from the as-grown film resemble very much those recorded from SrO, both in terms of the number of maxima and minima, their positions and the corresponding frequencies. However, the amplitudes of the oscillations are significantly smaller in the spectra of the as-grown film, the data becoming structureless beyond 6 $\AA^{-1}$. This strongly suggests the existence in the as-grown films of SrO-like domains having no long-range order what would explain why they have not been observed in the XRD data. The spectra recorded from the annealed film are totally consistent with the presence of a well crystalline SFO phase.

The Fe K-edge recorded from the different iron compounds is shown in Figure \ref{EXAFS}b. It is interesting to compare the data recorded from the as-grown film with those recorded from maghemite and hematite. Inspection of Figure \ref{EXAFS}b shows that the oscillations in the EXAFS of the as-grown film follow those of maghemite although with lower amplitude and being, in general, much structureless. However, they differ considerably from the data recorded from hematite. Again, this result is consistent with the Mössbauer data since it clearly points out the presence of poorly crystallized domains of maghemite in this film. As expected, the data recorded from the annealed film are fully consistent with well crystallized SFO.

Figures \ref{EXAFS}c and \ref{EXAFS}d collect the Sr- and Fe-K edge EXAFS transforms recorded from all these samples. We will not go into detail in the description of these since the conclusions which can be extracted are the same as those obtained from the EXAFS. The data recorded from the as-grown film shows the decreased intensity of the first coordination shell (both in the case of the strontium and iron data) and the absence of shells at longer distances, all the facts being indicative of the occurrence of a poorly crystalline material. This is not the case with the annealed film whose Fourier transforms are characteristic of SFO. 

\begin{figure}[htb]
\centerline{\includegraphics[width=0.5\textwidth]{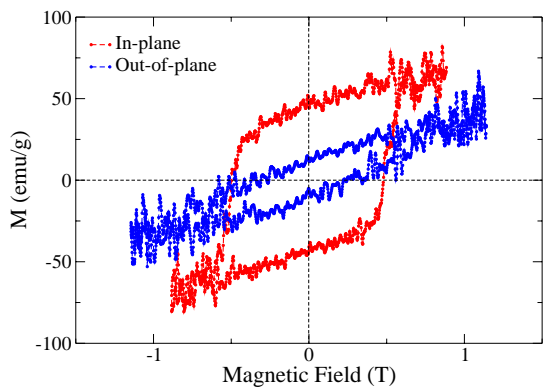}}
\caption{RT hysteresis loops registered from the SFO annealed film. The blue curve was recorded with a magnetic field applied normal to the sample while the red one was acquired with a magnetic field applied parallel to it.}
\label{VSMmicra}
\end{figure}

Finally, the magnetization curves of both films were recorded at RT under an applied external magnetic field of 1.25 T. For the as-grown sample, no magnetic easy axis or hysteresis loops were observed (not shown), in good agreement with the lack of magnetic interactions observed by Mössbauer spectroscopy at RT. Figure \ref{VSMmicra} shows the RT hysteresis loops recorded from the annealed film under an applied magnetic field parallel and normal to the sample. The results show clearly the existence of a magnetization easy axis within the plane of the sample. This is consistent with the preference of SFO to grow with the c-axis oriented parallel to the film plane, as indicated by XRD, and the fact that magnetization occurs in SFO along that axis. These data also confirm the orientation of the in-plane magnetization determined by Mössbauer spectroscopy at RT and 35 K (figure \ref{SFOmicramossbauerfinal}b). The saturation magnetization and coercive field determined with the in-plane configuration were 63 emu/g and 0.5 T, respectively. These values are within the range of magnetization values reported for this compound \cite{PullarProgMatSci2012,fernandez_topical_2020}. When the magnetization curve was recorded perpendicular to the plane of the sample, a different saturation magnetization value was found. This shows the need for a higher applied field to reach the threshold magnetization value observed previously in the in-plane measurement. The noisy signal is related both to the relatively small thickness of the films (a reduced amount of material results in a poorer signal) and the particular acquisition conditions of our homemade VSM device.

From the inspection of the results above, the precursor phase in the as-grown sample is quite isotropic. On one hand, the iron phase detected is locally cubic with a low magnetic anisotropy. On the other, the magnetization direction determined from Mössbauer at low temperature is compatible with a random orientation of the magnetization. However, once the film is annealed, it presents a clear uniaxial anisotropy with the easy axis within the film plane (see the VSM and Mössbauer results). We suggest that the origin of the in-plane magnetization is due to the film morphology that in the annealing stage promotes the growth of the SFO crystals with the c-axis in the in-plane direction.

\section{Conclusions}

In this work, we have confirmed that the annealing treatment is crucial for obtaining a genuine SFO film having the right composition and crystal structure. To this aim, two samples grown by RF magnetron sputtering with and without heating step were characterized by distinct diffraction and spectroscopy techniques. The XRD and Raman data showed the poor crystallinity of the as-grown sample as well as the possible presence of maghemite. Mössbauer spectroscopy corroborated the presence of nanometric $\gamma$-Fe$_{2}$O$_{3}$ domains in the non-annealed film while the XANES and EXAFS results revealed the existence of SrO-like domains having no long-range order. The EXAFS and XANES data taken at different incident angles of the incident X-ray beam in the as-grown films do not show significant differences and therefore do not support previous claims \cite{snyder_local_1995,snyder_local_1996, snyder_determination_1996} that the as-grown film acts as a “c-axis oriented” precursor of SFO. All the results gathered from the annealed film confirm that it corresponds to a canonical SFO phase which besides is highly textured from both the structural and magnetic points of view, having the c-axis and consequently, the easy magnetization axis parallel to the film plane.

\section{Acknowledgments}

This work is supported by the Spanish Ministry of Science and Innovation through Projects RTI2018-095303-B-C51, RTI2018-095303-B-C53, MAT2017-86450-C4-1-R, RTI2018-095303-A-C52, through the Ramon y Cajal Contract RYC-2017-23320 and by the Regional Government of Madrid through project S2018-NMT-4321 and by the European Commission through the H2020 Project no.720853 (AMPHIBIAN). We gratefully acknowledge the support from CMAM for the beamtime proposal with code STD009/20.
Part of these experiments were performed at the CLAESS-BL22 beamline at ALBA Synchrotron and in collaboration with ALBA staff.
A.S. acknowledges financial support from the Comunidad de Madrid for an “Atracción de Talento Investigador” contract No. 2017-t2/IND539.
We thank Dr S. Cortés for providing access to the Raman spectrometer and for his help in recording these data.

\section{CRediT authorship contribution statement}

G.D. Soria: conceptualization, investigation, formal analysis, software, visualization, writing-original draft; A. Serrano: investigation, resources, writing-review $\&$ editing; J.E. Prieto: methodology, investigation, software, writing-review $\&$ editing; A. Quesada: methodology, supervision, writing-review $\&$ editing, project administration, funding acquisition; G. Gorni: investigation, resources; J. de la Figuera: conceptualization, methodology, supervision, writing-review $\&$ editing, project administration, funding acquisition; J.F. Marco: conceptualization, investigation, formal analysis, visualization, writing-original draft.

\section{Additional Information}

\textbf{Competing Interests:} The authors declare that they have no known competing financial interests or personal relationships that could have appeared to influence the work reported in this paper.\\
\textbf{Data Availability:} Data will be made available on reasonable request.

\bibliography{SFOthinfilm}

\end{document}